\documentclass[a4paper,twocolumn,10pt,prb]{revtex4}
\usepackage {epsfig}
\newcommand{\degree}{\ensuremath{^\circ}}

\begin{document}

\bibliographystyle{apsrev}

\title{Exciton-photon coupling in a ZnSe based microcavity fabricated using epitaxial liftoff}
\author{A. Curran}
\author{J. K. Morrod}
\author{K. A. Prior}
\author{A. K. Kar}
\author{R. J. Warburton}
\affiliation{School of Engineering and Physical Sciences, Heriot-Watt University, Edinburgh EH14 4AS, UK}
\date{\today}

\begin{abstract}
We report the observation of strong exciton-photon coupling in a ZnSe based microcavity fabricated using epitaxial 
liftoff. Molecular beam epitaxial grown ZnSe/Zn$_{0.9}$Cd$_{0.1}$Se quantum wells with a one wavelength optical length 
at the exciton emission were transferred to a SiO$_2$/Ta$_2$O$_5$ mirror with a reflectance of $96\%$ to form finesse 
matched microcavities. Analysis of our angle resolved transmission spectra reveals key features of the strong coupling 
regime: anticrossing with a normal mode splitting of $23.6 meV$ at $20 K$; composite evolution of the lower and upper 
polaritons; and narrowing of the lower polariton linewidth near resonance. The heavy hole exciton oscillator strength 
per quantum well is also deduced to be $1.78 \times 10^{13} cm^{-2}$.
\end{abstract}

\maketitle


Light-matter interaction inside quantum microcavities (MC) containing one or more quantum wells has seen growing 
interest since the first experimental observation of the normal mode splitting of the coupled 
oscillators.\cite{prl_69_3314} The two coupled states, the lower and upper cavity-polaritons, are the result of the 
strong hybridization of the resonant cavity photon and quantum well exciton. The cavity-polariton lends itself to the 
possibility of a threshold-less laser,\cite{Imamoglu:polLaserTheory,LiSeDang:polLaser,Malpuech:pollasing} ultrafast 
micro-optical amplifiers,\cite{Savvidis:polariton,Saba:hotpolariton} and Bose-Einstein Condensation (BEC) in a solid 
state environment.\cite{Dang:BEC}

Much of the work done in this area has focused on III-V 
MCs\cite{prl_69_3314,prl_73_2043,Savvidis:polariton,Saba:hotpolariton} because of the high quality quantum wells and 
GaAs/AlAs distributed Bragg reflectors. However, the low binding energy of excitons found in typical III-V QWs ($5 - 10 
meV$) is not ideal. Saba \emph{et. al.}\cite{Saba:hotpolariton} show that the cut-off temperature for polariton 
parametric scattering, the proposed mechanism for a thresholdless laser, is determined by the exciton binding energy, 
and therefore room temperature operation is prohibited. Achieving BEC in MCs also requires sufficiently large binding 
energies such that the critical density can be reached before ionization. Wide bandgap II-VI based MCs are a promising 
alternative since the typical binding energy ($\sim 30 meV$) is comparable to the room temperature thermal 
energy.\cite{basicdata} Fabrication of II-VI MCs is a challenge with all growth techniques since suitable materials 
that satisfy the lattice matching criteria are in their infancy. To the best of our knowledge there are only a few 
demonstrations of II-VI based MCs, either involving growth-etch-growth processing\cite{Kelkar:26MC,pawlis:26MC} or the 
use of semiconductor mirrors.\cite{LiSeDang:polLaser,Feltin:GaN_MC,lohmeyer:26MC,Dang:BEC} Large light penetration 
depth due to the low refractive index contrast and difficulties of strain management are inherent disadvantages of the 
latter approach.

\begin{figure}[h]
\includegraphics[scale=0.24]{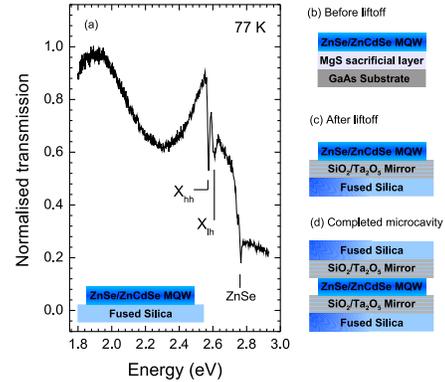}
\caption{(a) Normalised transmission of the ZnSe/Zn$_{0.9}$Cd$_{0.1}$Se quantum wells used as the active medium of the 
MC. Reference samples were lifted and transferred to a fused silica substrate. Inset: Schematic of the quantum wells 
transferred to a glass substrate. (b-d) Schematics of the sample at each stage of the MC fabrication (not to scale).}
\label{FIG1}
\end{figure}

We report here the successful fabrication and characterisation of ZnSe based MCs using a selective etching 
technique.\cite{Balocchi:lift} This new technology has allowed us to fabricate high quality ZnSe based MCs with 
dielectric mirrors. The large binding energies found in ZnSe based QWs combined with the high performance of dielectric 
mirrors makes our samples ideally suited to the study of parametric scattering and BEC.

The active region of the cavity was grown by MBE on high quality GaAs $n^+-$substrates following deposition of a $20 
nm$ ZnSe buffer layer. A sacrificial layer of MgS ($10 nm$) was grown\cite{Bradford:growth}, then the active region, 5 
ZnSe($8 nm$)/Zn$_{0.9}$Cd$_{0.1}$Se($8 nm$) quantum wells with ZnSe spacers ($\sim 53 nm$) top and bottom such that the 
thickness of the active layer was equal to one optical wavelength at the emission of the well. Rotation of the sample 
was stopped during the growth of the top ZnSe spacer so as to allow translational tuning of $\sim 10 meV mm^{-1}$ of 
the completed MC. Small $3 \times 3 mm$ samples from the wafer were prepared by cleaving and coated with wax. The MgS 
sacrificial layer was selectively etched in a solution of HCl acid to free the active layer from the GaAs substrate. 
The lifted layer, supported by the layer of wax, was then transferred to a SiO$_2$/Ta$_2$O$_5$ dielectric mirror. 
Details of the liftoff technique are reported elsewhere.\cite{Balocchi:lift} Our mirrors are specifically designed such 
that the cavity linewidth, $\gamma_{cav}$ and electric field profile were optimum for strong coupling with our quantum 
wells.\cite{coupling_criteria} The mirror transmission was measured to be $3.9 \%$ at $475 nm$ with a structure 
designated by $g(LH)^7La$ ($L:SiO_2, n = 1.46$, $H:Ta_2O_5, n= 2.1$ and $a:air, n=1.0 $). Finally a second dielectric 
mirror was mechanically held against the active layer forming a MC of cavity length $L_{cav} \sim 170 
nm$\cite{epithick} and refractive index $n_{cav} = 2.78$. The completed MCs were transferred to a continuous flow 
cryostat and characterized at $20 K$. All our fabrication steps were carried out in an air filtered environment, 
reducing the possibility of dust particles becoming trapped within the MC.

Reference samples were lifted and transferred to fused silica substrates and characterised by measuring their 
transmission at $77 K$. White-light continuum generated with $fs$ pulses from a Ti:Saphire laser was focused to a spot 
size of $~20 \mu m$ on the sample. The sample was normal to the optical axis. The transmitted light was collected with 
an angular resolution of $0.2 \degree$ at normal incidence by a $500 \mu m$ multi-mode fibre bundle. The collected 
light was dispersed with a spectrometer with a resolution of $0.1 nm$ and imaged onto a liquid nitrogen cooled CCD 
camera. Figure \ref{FIG1} shows the transmission of $5$ ZnSe/Zn$_{0.9}$Cd$_{0.1}$Se quantum wells lifted and 
transferred to a fused silica substrate. Two main optical features are the heavy-hole exciton $X_{hh} = 2.575 eV$ and 
light-hole exciton $X_{lh} = 2.611 eV$ transitions. A low energy oscillation is clear and is attributed to interference 
within the overall epitaxial layer. High energy features around $2.77 eV$ correspond to bulk transitions in the ZnSe 
barriers.

\begin{figure}[h]
\includegraphics[scale=0.6]{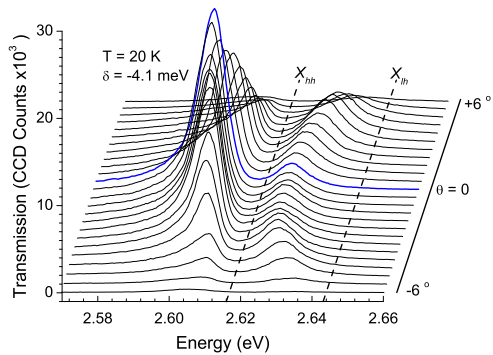}
\caption{Angle resolved transmission spectra taken at spatial points in the first Fourier plane of the collection lens. 
The sample was detuned by $-4.1 meV$ at $\theta=0 \degree$ and scanned through $\pm 6 \degree$. The dashed lines mark 
the uncoupled heavy hole, $X_{hh}$ and light hole exciton, $X_{lh}$ transitions. The normal incidence spectrum is 
highlighted. Spectra are offset for clarity.}
\label{FIG2}
\end{figure}

The definitive experiment when searching for cavity-polaritons is the measurement of the optical angular dispersion. In 
the weak coupling regime the two states disperse independently. In the strong coupling regime the observed dispersion 
curves are distinctly altered on and near resonance, where they anticross with a normal mode splitting (Rabi 
splitting), typically a few $meV$.

Completed microcavities were characterised by measuring white light transmission at $20 K$. Two identical lenses, NA 
matched to the cryostat, were used to focus and collect light at the sample. Fully illuminating the objective lens, 
white light was focused at the sample and the transmitted light was collimated to a diameter of $33 mm$, where a fibred 
coupled spectrometer was used to scan through the collimated light and thus record a series of transmission spectra. In 
this way lateral displacement is related to angular dispersion inside the MC. Taking into account the glass substrate 
of the mirrors the range of angle at the MC surface was $\pm 6 \degree$. Each spectrum was taken at different points 
through the centre of the collimated beam with increments of $1 mm$. Since the in-plane wavevector inside the cavity, 
$k_{||}$ is related to the angle of incident light by $k_{||} = \frac{\omega_{cav}}{c} sin (\theta)$ we can measure the 
dispersion curves of our samples in $k$-space between $\pm 1.39 \times10^6 m^{-1}$. Figure \ref{FIG2} shows a series of 
spectra taken for one of our samples detuned to $\delta = -4.1 meV$ relative to the $X_{hh}$ transition at normal 
incidence. The spectra are offset with the lowest spectrum corresponding to $k_{||}=-1.39 \times10^6 m^{-1}$ 
($-6\degree$) increasing to $k_{||}=1.39 \times10^6 m^{-1}$ ($+6\degree$). The uncoupled $X_{hh}$ and $X_{lh}$ 
transition are indicated by the dotted lines. Two optical features clearly anticross on either side of the $X_{hh}$ 
transition, the lower (LP) and upper polariton (UP), a clear demonstration of strong coupling.

\begin{figure}[h]
\includegraphics[scale=0.85]{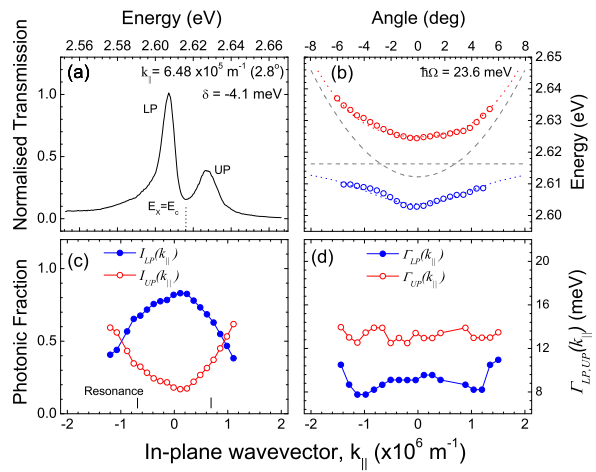}
\caption{(a) Transmission spectrum on resonance for $\delta=-4.1 meV$ and $k_{||}= 6.48 \times 10^5 m^{-1}$ 
($2.8\degree$). (b) Energy peak values from figure \ref{FIG2} (open circles). Resonance occurs at $k_{||} = \pm 6.48 
\times 10^5$ ($\pm 2.8\degree$) with $\hbar \Omega = 23.6 meV$. The polariton dispersion curves are compared to the 
model discussed in the text (dotted lines). Bare cavity and $X_{hh}$ dispersion curves are also shown (dashed lines). 
(c) Integrated transmission intensity for the lower and upper polariton states. (d) Measured polariton linewidths, 
$\Gamma_{LP,UP}(k_{||})$.}
\label{FIG3}
\end{figure}

Figure \ref{FIG3}(a) shows the resonant transmission spectrum for the same detuning as figure \ref{FIG2}. At this
point on the dispersion curve, $k_{||}= 6.48 \times 10^5 m^{-1}$ ($2.8\degree$), the composite states are in resonance 
and the energy separation between the two modes is a minimum. The peak energy values for the LP and UP are extracted 
from figure \ref{FIG2} and presented in figure \ref{FIG3}(b) (open circles) as a function of $k_{||}$. Treating the 
composite states of the polariton as coupled oscillators the eigenvalues of the coupling Hamiltonian were used to 
calculate the energy dispersion of the LP and UP taking the Rabi splitting, $\hbar \Omega$ as fit 
parameter.\cite{Savona:poltheory} The coupled and bare states are displayed as dotted and dashed lines respectively and 
give good agreement with the experimental data. The point of resonance is found to occur at $\pm 2.8\degree$ with 
$\hbar \Omega = 23.6 meV$. The heavy hole exciton oscillator strength, $f_{X_{hh}}$ can be determined from the 
expression,\cite{Savona:poltheory} $f_{X_{hh}} = \frac{(\hbar \Omega)^2 n_c^2 L_{e} \epsilon_o m_o}{2 e^2 \hbar^2 
N_{qw}}$, where the number of quantum wells, $N_{qw}$, the cavity refractive index, $n_{c}$ and effective cavity 
length, $L_{e}$ are $5$, $2.78$ and $571 nm$ respectively. $m_o$ is the free electron mass. For our microcavity we 
deduce an oscillator strength per quantum well, $f_{X_{hh}}=1.78 \times 10^{13} cm^{-2}$. This is similiar to 
$f_{X_{hh}} \sim 1 \times 10^{13} cm^{-2}$ for MCs studied by Kelkar \emph{et. al.}.\cite{Kelkar:26MC}

The polariton state transmission amplitude is a measure of the polariton composite ratio.\cite{Hopfield:fraction} In 
Figure \ref{FIG3}(c) we present the normalized integrated intensities, $I_{LP,UP}(k_{||})$ of the LP and UP branch 
giving a proportional indication of the evolution of the photonic fraction of the polariton. We have defined 
$I_{LP,UP}(k_{||}) = \frac{I_{lp,up}(k_{||})}{I_{lp}(k_{||})+I_{up}(k_{||})}$ such that 
$I_{LP}(k_{||})+I_{UP}(k_{||})=1$. In the angular range of $-4 \degree \leq 0 \leq + 4 \degree$ the LP is predominantly 
photon-like, indicated by the increase in the LP intensity. Beyond this range the LP weakens as it becomes 
progressively more excitonic and $E_{LP}(k_{||}) \rightarrow E_{X_{hh}}$. The UP behaves inversely. We also note that 
on resonance, figure \ref{FIG3}(a), $I_{lp}(k_{||}) = 2.6 \times I_{up}(k_{||})$. $I_{lp}(k_{||}) = I_{lp}(k_{||})$ 
holds at larger $k_{||}$ beyond resonance. This is also the case in the MC studied by Kelkar \emph{et. 
al.}.\cite{Kelkar:26MC} The measured linewidths, $\Gamma_{LP,UP}(k_{||})$ of the LP and UP are presented in figure 
\ref{FIG3} (d). Narrowing of the LP linewidth is clearly evident and occurs at $\sim k_{||}= \pm 1 \times 10^6 m^{-1}$ 
($\sim \pm 3 \degree$).\cite{Kavokin:linewidth} This is evidence that the extended nature of the polariton averages out 
some of the inhomogeneous broadening in the quantum well exciton.

In conclusion, we have shown that high quality ZnSe based MCs can be fabricated with the use of an epitaxial liftoff 
technique.\cite{Balocchi:lift} We have successfully demonstrated three key features of the strong coupling regime: 
anticrossing of the cavity polariton with a normal mode splitting of $23.6 meV$; composite fractional evolution of the 
LP and UP; and narrowing of the LP linewidth near resonance. We also deduce the heavy hole exciton oscillator strength 
to be $f_{X_{hh}}=1.78 \times 10^{13} cm^{-2}$. Our fabrication technique has eliminated the need for complicated 
growth-etch-growth processing whilst still employing the advantages of large exciton binding energies found in wide 
bandgap II-VI materials and high reflectivity offered by dielectric mirrors. Additionally, the epitaxial liftoff 
technology can potentially be developed to fabricate MCs with MgS/ZnSe quantum wells with an even higher exciton 
binding energies, $\sim 40 meV$.\cite{bernard:Eb} The combination of prolonged photonic lifetimes, due to the 
achievable high finesse, and the large excitonic oscillator strength are ideally suited to nonlinear phenomena such as 
parametric scattering and BEC.

\noindent This work was funded by EPSRC, UK.


\begin{thebibliography}{21}
\expandafter\ifx\csname natexlab\endcsname\relax\def\natexlab#1{#1}\fi
\expandafter\ifx\csname bibnamefont\endcsname\relax
  \def\bibnamefont#1{#1}\fi
\expandafter\ifx\csname bibfnamefont\endcsname\relax
  \def\bibfnamefont#1{#1}\fi
\expandafter\ifx\csname citenamefont\endcsname\relax
  \def\citenamefont#1{#1}\fi
\expandafter\ifx\csname url\endcsname\relax
  \def\url#1{\texttt{#1}}\fi
\expandafter\ifx\csname urlprefix\endcsname\relax\def\urlprefix{URL }\fi
\providecommand{\bibinfo}[2]{#2}
\providecommand{\eprint}[2][]{\url{#2}}

\bibitem[{\citenamefont{Weisbuch et~al.}(1992)\citenamefont{Weisbuch, Nishioka,
  Ishikawa, and Arakawa}}]{prl_69_3314}
\bibinfo{author}{\bibfnamefont{C.}~\bibnamefont{Weisbuch}},
  \bibinfo{author}{\bibfnamefont{M.}~\bibnamefont{Nishioka}},
  \bibinfo{author}{\bibfnamefont{A.}~\bibnamefont{Ishikawa}}, \bibnamefont{and}
  \bibinfo{author}{\bibfnamefont{Y.}~\bibnamefont{Arakawa}},
  \bibinfo{journal}{Phys. Rev. Lett.} \textbf{\bibinfo{volume}{69}},
  \bibinfo{pages}{3314} (\bibinfo{year}{1992}).

\bibitem[{\citenamefont{Imamo\={g}lu et~al.}(1996)\citenamefont{Imamo\={g}lu,
  Ram, Pau, and Yamamoto}}]{Imamoglu:polLaserTheory}
\bibinfo{author}{\bibfnamefont{A.}~\bibnamefont{Imamo\={g}lu}},
  \bibinfo{author}{\bibfnamefont{R.~J.} \bibnamefont{Ram}},
  \bibinfo{author}{\bibfnamefont{S.}~\bibnamefont{Pau}}, \bibnamefont{and}
  \bibinfo{author}{\bibfnamefont{Y.}~\bibnamefont{Yamamoto}},
  \bibinfo{journal}{Phys. Rev. A} \textbf{\bibinfo{volume}{53}},
  \bibinfo{pages}{4250} (\bibinfo{year}{1996}).

\bibitem[{\citenamefont{Dang et~al.}(1998)\citenamefont{Dang, Heger, Andr\'{e},
  B\oe{}uf, and Romestain}}]{LiSeDang:polLaser}
\bibinfo{author}{\bibfnamefont{L.~S.} \bibnamefont{Dang}},
  \bibinfo{author}{\bibfnamefont{D.}~\bibnamefont{Heger}},
  \bibinfo{author}{\bibfnamefont{R.}~\bibnamefont{Andr\'{e}}},
  \bibinfo{author}{\bibfnamefont{F.}~\bibnamefont{B\oe{}uf}}, \bibnamefont{and}
  \bibinfo{author}{\bibfnamefont{R.}~\bibnamefont{Romestain}},
  \bibinfo{journal}{Phys. Rev. Lett.} \textbf{\bibinfo{volume}{81}},
  \bibinfo{pages}{3920} (\bibinfo{year}{1998}).

\bibitem[{\citenamefont{Malpuech et~al.}(2002)\citenamefont{Malpuech, Carlo,
  Kavokin, Baumberg, Zamfirescu, and Lugli}}]{Malpuech:pollasing}
\bibinfo{author}{\bibfnamefont{G.}~\bibnamefont{Malpuech}},
  \bibinfo{author}{\bibfnamefont{A.~D.} \bibnamefont{Carlo}},
  \bibinfo{author}{\bibfnamefont{A.}~\bibnamefont{Kavokin}},
  \bibinfo{author}{\bibfnamefont{J.~J.} \bibnamefont{Baumberg}},
  \bibinfo{author}{\bibfnamefont{M.}~\bibnamefont{Zamfirescu}},
  \bibnamefont{and} \bibinfo{author}{\bibfnamefont{P.}~\bibnamefont{Lugli}},
  \bibinfo{journal}{Appl. Phys. Lett.} \textbf{\bibinfo{volume}{81}},
  \bibinfo{pages}{412} (\bibinfo{year}{2002}).

\bibitem[{\citenamefont{Savvidis et~al.}(2000)\citenamefont{Savvidis, Baumberg,
  Stevenson, Skolnick, Whittaker, and Roberts}}]{Savvidis:polariton}
\bibinfo{author}{\bibfnamefont{P.~G.} \bibnamefont{Savvidis}},
  \bibinfo{author}{\bibfnamefont{J.~J.} \bibnamefont{Baumberg}},
  \bibinfo{author}{\bibfnamefont{R.~M.} \bibnamefont{Stevenson}},
  \bibinfo{author}{\bibfnamefont{M.~S.} \bibnamefont{Skolnick}},
  \bibinfo{author}{\bibfnamefont{D.~M.} \bibnamefont{Whittaker}},
  \bibnamefont{and} \bibinfo{author}{\bibfnamefont{J.~S.}
  \bibnamefont{Roberts}}, \bibinfo{journal}{Phys. Rev. Lett.}
  \textbf{\bibinfo{volume}{84}}, \bibinfo{pages}{1547} (\bibinfo{year}{2000}).

\bibitem[{\citenamefont{Saba et~al.}(2001)\citenamefont{Saba, Ciuti, Bloch,
  Thierry-Mieg, Andre, Dang, Kundermann, Mura, {G. Bongiovanni, J. L. Staehli},
  and Deveaud}}]{Saba:hotpolariton}
\bibinfo{author}{\bibfnamefont{M.}~\bibnamefont{Saba}},
  \bibinfo{author}{\bibfnamefont{C.}~\bibnamefont{Ciuti}},
  \bibinfo{author}{\bibfnamefont{J.}~\bibnamefont{Bloch}},
  \bibinfo{author}{\bibfnamefont{V.}~\bibnamefont{Thierry-Mieg}},
  \bibinfo{author}{\bibfnamefont{R.}~\bibnamefont{Andre}},
  \bibinfo{author}{\bibfnamefont{L.~S.} \bibnamefont{Dang}},
  \bibinfo{author}{\bibfnamefont{S.}~\bibnamefont{Kundermann}},
  \bibinfo{author}{\bibfnamefont{A.}~\bibnamefont{Mura}},
  \bibinfo{author}{\bibnamefont{{G. Bongiovanni, J. L. Staehli}}},
  \bibnamefont{and} \bibinfo{author}{\bibfnamefont{B.}~\bibnamefont{Deveaud}},
  \bibinfo{journal}{Nature} \textbf{\bibinfo{volume}{414}},
  \bibinfo{pages}{731} (\bibinfo{year}{2001}).

\bibitem[{\citenamefont{Kasprzak et~al.}(2006)\citenamefont{Kasprzak, Richard,
  Kundermann, Baas, Jeambrun, Keeling, M.Marchetti, Szymanska, Andre, Staehli
  et~al.}}]{Dang:BEC}
\bibinfo{author}{\bibfnamefont{J.}~\bibnamefont{Kasprzak}},
  \bibinfo{author}{\bibfnamefont{M.}~\bibnamefont{Richard}},
  \bibinfo{author}{\bibfnamefont{S.}~\bibnamefont{Kundermann}},
  \bibinfo{author}{\bibfnamefont{A.}~\bibnamefont{Baas}},
  \bibinfo{author}{\bibfnamefont{P.}~\bibnamefont{Jeambrun}},
  \bibinfo{author}{\bibfnamefont{J.~M.~J.} \bibnamefont{Keeling}},
  \bibinfo{author}{\bibfnamefont{F.}~\bibnamefont{M.Marchetti}},
  \bibinfo{author}{\bibfnamefont{M.~H.} \bibnamefont{Szymanska}},
  \bibinfo{author}{\bibfnamefont{R.}~\bibnamefont{Andre}},
  \bibinfo{author}{\bibfnamefont{J.~L.} \bibnamefont{Staehli}},
  \bibnamefont{et~al.}, \bibinfo{journal}{Nature}
  \textbf{\bibinfo{volume}{443}}, \bibinfo{pages}{409} (\bibinfo{year}{2006}).

\bibitem[{\citenamefont{Houdr{\'e} et~al.}(1994)\citenamefont{Houdr{\'e},
  Weisbuch, Stanley, Oesterle, Pellandini, and Ilegems}}]{prl_73_2043}
\bibinfo{author}{\bibfnamefont{R.}~\bibnamefont{Houdr{\'e}}},
  \bibinfo{author}{\bibfnamefont{C.}~\bibnamefont{Weisbuch}},
  \bibinfo{author}{\bibfnamefont{R.~P.} \bibnamefont{Stanley}},
  \bibinfo{author}{\bibfnamefont{U.}~\bibnamefont{Oesterle}},
  \bibinfo{author}{\bibfnamefont{P.}~\bibnamefont{Pellandini}},
  \bibnamefont{and} \bibinfo{author}{\bibfnamefont{M.}~\bibnamefont{Ilegems}},
  \bibinfo{journal}{Phys. Rev. Lett.} \textbf{\bibinfo{volume}{73}},
  \bibinfo{pages}{2043} (\bibinfo{year}{1994}).

\bibitem[{\citenamefont{Madelung}(1996)}]{basicdata}
\bibinfo{editor}{\bibfnamefont{O.}~\bibnamefont{Madelung}}, ed.,
  \emph{\bibinfo{title}{Semiconductors - Basic Data}}
  (\bibinfo{publisher}{Springer}, \bibinfo{year}{1996}), \bibinfo{edition}{2nd}
  ed.

\bibitem[{\citenamefont{Kelkar et~al.}(1995)\citenamefont{Kelkar, Kozlov, Jeon,
  Nurmikko, Chu, Grillo, Han, Hua, and Gunshor}}]{Kelkar:26MC}
\bibinfo{author}{\bibfnamefont{P.}~\bibnamefont{Kelkar}},
  \bibinfo{author}{\bibfnamefont{V.}~\bibnamefont{Kozlov}},
  \bibinfo{author}{\bibfnamefont{H.}~\bibnamefont{Jeon}},
  \bibinfo{author}{\bibfnamefont{A.~V.} \bibnamefont{Nurmikko}},
  \bibinfo{author}{\bibfnamefont{C.~C.} \bibnamefont{Chu}},
  \bibinfo{author}{\bibfnamefont{D.~C.} \bibnamefont{Grillo}},
  \bibinfo{author}{\bibfnamefont{J.}~\bibnamefont{Han}},
  \bibinfo{author}{\bibfnamefont{C.~G.} \bibnamefont{Hua}}, \bibnamefont{and}
  \bibinfo{author}{\bibfnamefont{R.~L.} \bibnamefont{Gunshor}},
  \bibinfo{journal}{Phys. Rev. B} \textbf{\bibinfo{volume}{52}},
  \bibinfo{pages}{R5491} (\bibinfo{year}{1995}).

\bibitem[{\citenamefont{Pawlis et~al.}(2002)\citenamefont{Pawlis, Khartchenko,
  Husberg, As, Lischka, and Schikora}}]{pawlis:26MC}
\bibinfo{author}{\bibfnamefont{A.}~\bibnamefont{Pawlis}},
  \bibinfo{author}{\bibfnamefont{A.}~\bibnamefont{Khartchenko}},
  \bibinfo{author}{\bibfnamefont{O.}~\bibnamefont{Husberg}},
  \bibinfo{author}{\bibfnamefont{D.~J.} \bibnamefont{As}},
  \bibinfo{author}{\bibfnamefont{K.}~\bibnamefont{Lischka}}, \bibnamefont{and}
  \bibinfo{author}{\bibfnamefont{D.}~\bibnamefont{Schikora}},
  \bibinfo{journal}{Solid State Comm.} \textbf{\bibinfo{volume}{123}},
  \bibinfo{pages}{235} (\bibinfo{year}{2002}).

\bibitem[{\citenamefont{Feltin et~al.}(2006)\citenamefont{Feltin, Christmann,
  Butt\'{e}, Carlin, Mosca, and Grandjean}}]{Feltin:GaN_MC}
\bibinfo{author}{\bibfnamefont{E.}~\bibnamefont{Feltin}},
  \bibinfo{author}{\bibfnamefont{G.}~\bibnamefont{Christmann}},
  \bibinfo{author}{\bibfnamefont{R.}~\bibnamefont{Butt\'{e}}},
  \bibinfo{author}{\bibfnamefont{J.}~\bibnamefont{Carlin}},
  \bibinfo{author}{\bibfnamefont{M.}~\bibnamefont{Mosca}}, \bibnamefont{and}
  \bibinfo{author}{\bibfnamefont{N.}~\bibnamefont{Grandjean}},
  \bibinfo{journal}{Appl. Phys. Lett.} \textbf{\bibinfo{volume}{89}},
  \bibinfo{pages}{071107} (\bibinfo{year}{2006}).

\bibitem[{\citenamefont{Lohmeyer et~al.}(2006)\citenamefont{Lohmeyer, Sebald,
  Kruse, Kr{\"o}ger, Gutowski, Hommel, Wiersig, Baer, and
  Jahnke}}]{lohmeyer:26MC}
\bibinfo{author}{\bibfnamefont{H.}~\bibnamefont{Lohmeyer}},
  \bibinfo{author}{\bibfnamefont{K.}~\bibnamefont{Sebald}},
  \bibinfo{author}{\bibfnamefont{C.}~\bibnamefont{Kruse}},
  \bibinfo{author}{\bibfnamefont{R.}~\bibnamefont{Kr{\"o}ger}},
  \bibinfo{author}{\bibfnamefont{J.}~\bibnamefont{Gutowski}},
  \bibinfo{author}{\bibfnamefont{D.}~\bibnamefont{Hommel}},
  \bibinfo{author}{\bibfnamefont{J.}~\bibnamefont{Wiersig}},
  \bibinfo{author}{\bibfnamefont{N.}~\bibnamefont{Baer}}, \bibnamefont{and}
  \bibinfo{author}{\bibfnamefont{F.}~\bibnamefont{Jahnke}},
  \bibinfo{journal}{Appl. Phys. Lett.} \textbf{\bibinfo{volume}{88}},
  \bibinfo{pages}{51101} (\bibinfo{year}{2006}).

\bibitem[{\citenamefont{Balocchi et~al.}(2005)\citenamefont{Balocchi, Curran,
  Graham, Bradford, Prior, and Warburton}}]{Balocchi:lift}
\bibinfo{author}{\bibfnamefont{A.}~\bibnamefont{Balocchi}},
  \bibinfo{author}{\bibfnamefont{A.}~\bibnamefont{Curran}},
  \bibinfo{author}{\bibfnamefont{T.~C.~M.} \bibnamefont{Graham}},
  \bibinfo{author}{\bibfnamefont{C.}~\bibnamefont{Bradford}},
  \bibinfo{author}{\bibfnamefont{K.~A.} \bibnamefont{Prior}}, \bibnamefont{and}
  \bibinfo{author}{\bibfnamefont{R.~J.} \bibnamefont{Warburton}},
  \bibinfo{journal}{Appl. Phys. Lett.} \textbf{\bibinfo{volume}{86}},
  \bibinfo{pages}{011916} (\bibinfo{year}{2005}).

\bibitem[{\citenamefont{Bradford et~al.}(2001)\citenamefont{Bradford,
  O'Donnell, Urbaszek, Balocchi, Morhain, Prior, and
  Cavenett}}]{Bradford:growth}
\bibinfo{author}{\bibfnamefont{C.}~\bibnamefont{Bradford}},
  \bibinfo{author}{\bibfnamefont{C.~B.} \bibnamefont{O'Donnell}},
  \bibinfo{author}{\bibfnamefont{B.}~\bibnamefont{Urbaszek}},
  \bibinfo{author}{\bibfnamefont{A.}~\bibnamefont{Balocchi}},
  \bibinfo{author}{\bibfnamefont{C.}~\bibnamefont{Morhain}},
  \bibinfo{author}{\bibfnamefont{K.~A.} \bibnamefont{Prior}}, \bibnamefont{and}
  \bibinfo{author}{\bibfnamefont{B.~C.} \bibnamefont{Cavenett}},
  \bibinfo{journal}{J. Cryst. Growth} \textbf{\bibinfo{volume}{227}},
  \bibinfo{pages}{634} (\bibinfo{year}{2001}).

\bibitem[{cou()}]{coupling_criteria}
\bibinfo{note}{The bare exciton linewidth, $\gamma_{X_{hh}}= 9.2 meV$ was
  matched to $\gamma_{cav} \sim 15 meV$ satisfying, $\Omega^2 \gg
  (\gamma_{X_{hh}}-\gamma_{cav})^2$. The stacking order of the
  $\frac{\lambda}{4}$ layers was such that the antinode of the electric field
  inside the cavity was positioned at the quantum wells.}

\bibitem[{epi()}]{epithick}
\bibinfo{note}{The epitaxial layer thickness deduced from figure 1 was used in the coupled
  oscillator model in figure 3(b).}

\bibitem[{\citenamefont{Savona et~al.}(1995)\citenamefont{Savona, Andreani,
  Schwendimann, and Quattropani}}]{Savona:poltheory}
\bibinfo{author}{\bibfnamefont{V.}~\bibnamefont{Savona}},
  \bibinfo{author}{\bibfnamefont{L.~C.} \bibnamefont{Andreani}},
  \bibinfo{author}{\bibfnamefont{P.}~\bibnamefont{Schwendimann}},
  \bibnamefont{and}
  \bibinfo{author}{\bibfnamefont{A.}~\bibnamefont{Quattropani}},
  \bibinfo{journal}{Solid State Comm.} \textbf{\bibinfo{volume}{93}},
  \bibinfo{pages}{733} (\bibinfo{year}{1995}).

\bibitem[{\citenamefont{Hopfield}(1958)}]{Hopfield:fraction}
\bibinfo{author}{\bibfnamefont{J.~J.} \bibnamefont{Hopfield}},
  \bibinfo{journal}{Phys. Rev.} \textbf{\bibinfo{volume}{112}},
  \bibinfo{pages}{1555} (\bibinfo{year}{1958}).

\bibitem[{\citenamefont{Kavokin}(1998)}]{Kavokin:linewidth}
\bibinfo{author}{\bibfnamefont{A.~V.} \bibnamefont{Kavokin}},
  \bibinfo{journal}{Phys. Rev. B} \textbf{\bibinfo{volume}{57}},
  \bibinfo{pages}{3757} (\bibinfo{year}{1998}).

\bibitem[{\citenamefont{Urbaszek et~al.}(2000)\citenamefont{Urbaszek, Balocchi,
  Bradford, Morhain, O'Donnell, Prior, and Cavenett}}]{bernard:Eb}
\bibinfo{author}{\bibfnamefont{B.}~\bibnamefont{Urbaszek}},
  \bibinfo{author}{\bibfnamefont{A.}~\bibnamefont{Balocchi}},
  \bibinfo{author}{\bibfnamefont{C.}~\bibnamefont{Bradford}},
  \bibinfo{author}{\bibfnamefont{C.}~\bibnamefont{Morhain}},
  \bibinfo{author}{\bibfnamefont{C.~B.} \bibnamefont{O'Donnell}},
  \bibinfo{author}{\bibfnamefont{K.~A.} \bibnamefont{Prior}}, \bibnamefont{and}
  \bibinfo{author}{\bibfnamefont{B.~C.} \bibnamefont{Cavenett}},
  \bibinfo{journal}{Appl. Phys. Lett.} \textbf{\bibinfo{volume}{77}},
  \bibinfo{pages}{3755} (\bibinfo{year}{2000}).

\end{thebibliography}

\end{document}